\documentclass[usenatbib,usegraphicx]{mn2e}

\usepackage{times}
\usepackage{amssymb}

\newcommand{\aap}{A\&A}
\newcommand{\aaps}{A\&AS}

\newcommand{\aj}{AJ}
\newcommand{\apj}{ApJ}
\newcommand{\apjl}{ApJ}
\newcommand{\apjs}{ApJS}
\newcommand{\mnras}{MNRAS}
\newcommand{\pasp}{PASP}
\newcommand{\apss}{Ap\&SS}
\newcommand{\nat}{Nature}

\newcommand{\uno}{PKS\,0048$-$097}%
\newcommand{\dos}{PKS\,0754$+$100}%
\newcommand{\tres}{PKS\,1510$-$089}%

\defcitealias{HWM88}{HWM88}
\begin{document}
\title[Extremely violent microvariability in blazars]{Extremely violent optical
microvariability in blazars: fact or fiction?\thanks{Based on observations
collected at the Complejo Astron\'omico El Le\-on\-ci\-to, operated under
agreement among CONICET, UNLP, UNC, and UNSJ, Argentina.}}

\author[Cellone et al.]{Sergio A. Cellone$^{1}$, Gustavo E. Romero$^{1,2}$,
and Anabella T.  Araudo$^{1,2}$\\
$^1$Facultad de Ciencias Astron\'omicas y Geof\'\i sicas, Universidad Nacional
de La Plata, Paseo del Bosque, B1900FWA La Plata, Argentina\\
$^2$Inst.\ Argentino de Radioastronom\'{\i}a, C.C.5, 1894 Villa
Elisa,  Buenos Aires, Argentina
}

\maketitle

\begin{abstract}
Variability amplitudes larger than 1 magnitude over time-scales of a few tens
of minutes have recently been reported in the optical light-curves of
several blazars. In order to independently verify the real occurrence of
such extremely violent events, we undertook an observational study of a
selected sample of three blazars: \uno, \dos, and \tres.  Possible
systematic error sources during data acquisition and reduction were
carefully evaluated.  We indeed found flux variability at intra-night
time-scales in all three sources, although no extremely violent behaviour, as
reported by other authors, was detected. We show that an incorrect choice of
the stars used for differential photometry will, under fairly normal
conditions, lead to spurious variability with large amplitudes on short
time-scales. Wrong results of this kind can be avoided with the use of
simple error-control techniques.
\end{abstract}


\section{Introduction}
 \label{s_intro}

One distinctive characteristic of active galactic nuclei (AGN) is the fact
that their energy flux varies along the whole electromagnetic spectrum,
spanning a wide range of time-scales. Indeed, flux variability is an often
used criterion for AGN detection \citep[e.g.:][]{M04}. Within the widely
accepted canonical model, i.e., super-massive black-hole (SMBH) +
accretion-disk (AD) + dusty torus (DT) + relativistic jets, different
regions of the active nucleus are thought to contribute to the power emitted
at different frequencies along the spectral energy distribution (SED). On
the other hand, variability time-scales strongly constrain the sizes of the
emitting regions, through light-travel arguments.  Since, except for the
Mpc-scale radio-jets and lobes, the other components remain spatially
unresolved to astronomical observations in the vast majority of AGNs,
flux variability studies are thus a powerful tool to probe those innermost
regions.

While most AGNs are variable on time-scales of a few years, blazars, i.e.,
the subclass comprising BL\,Lac objects and flat-spectrum radio-quasars
(FSRQ), display both the largest amplitudes and shorter time-scales.  At
optical wavelengths, long-term variations, with amplitudes of $\sim 2$ to
$\sim 5$ mag along a few years have been found through extensive monitoring
in several objects, such as GC\,0109+224 \citep{CTR03}, AO\,0235+64
\citep{RVI05}, and OJ\,287 \citep{QT03}.
 Optical fluctuations on time-scales spanning several days are also usually
observed, although with smaller amplitudes (e.g., $\sim 1$ mag in 20
days for PKS\,2005$-$489, \citealp{DATB04}; $\sim 1.7$ mag in 10
days for 3C 454.3, \citealp{FCM06}).

Although photometric changes for blazars on very short time-scales were
reported more than 30 years ago \citep[$\Delta m \simeq 0.3$ mag in less
than 24 hs for BL\,Lac,][]{R70}, this phenomenon remained unrecognized by
most astronomers until the advent of CCD detectors, when the existence of
the now called intra-night optical variability (INOV) or
\emph{microvariability} was firmly established \citep*{MCG89, CMG90, CMNS91,
CMNG92}. Soon it became clear that several sources could experience
remarkably large intra-night fluctuations, amounting to several tenths of a
magnitude in a few hours. This is the case, among others, for
PKS\,0537$-$441 \citep*{TBB86, HW96, RCC00aj}, 
3C\,371 \citep{CNM98}, and AO\,0235+64 \citep{RMSS96, HW96, NM96}.
For this last object, \citet*{RCC00aa} found changes up to 0.5 mag within one
night and $\sim 1.2$ mag between consecutive nights, through well-sampled
$V$ and $R$ band light-curves. This was one of the most violent variability
events ever observed at optical wavelengths in any blazar.
 The statistical incidence of microvariability in different classes of AGNs
has been studied by several authors \citep*{RCC99, GSSW03, SSGW04, SGG05}.

The observed optical flux in blazars originates in a part of the AD and the
inner (pc-scale) portions of the jets.  Disregarding gravitational
micro-lensing effects, which probably apply to a small subset of particular
sources \citep[e.g.:][]{NCP96}, three broad classes of intrinsic models may
explain optical microvariability: a) hot-spot models, involving
instabilities in the AD \citep{MW93}, b) shock-in-jet models, based on a
relativistic shock-front interacting with inhomogeneities or bends in the
jet \citep*{MG85, R95, RCV95}, and c) geometrical models where the jet
changes its orientation relative to the observer, thus changing its Doppler
factor \citep{M92, GW92}. Violent phenomena as those above reported,
probably rule out models based on AD instabilities, needing, instead,
relativistic beaming with high Doppler factors.

In recent years, several papers claimed the repeated detection of extremely
violent variability events in some blazars, with amplitudes $\Delta m
\gtrsim 1$ mag in a few tens of minutes \citep{BXL98, XLZBL99, XLB01, XLZ02,
XZD02, XZL04, DXL01}. For example, these authors reported a 2 mag variation
in $\sim40$ min for the highly polarized quasar (HPQ) PKS\,1510$-$089; they
also showed several sudden ($\Delta t \lesssim 15$ min) ``dips'' of $\sim
0.9$ mag in the light-curve of the EGRET blazar OJ\,248 (0827+243).  If
confirmed, these extremely violent phenomena would require a complete
reassessment of the mechanisms which are thought to be responsible of the
energy generation in AGNs. As an illustration, let us mention that optical
variability time-scales of a few tens of minutes would imply emitting regions
smaller than the Schwarzschild radius for certain objects.

In contradiction to \citeauthor{XLZBL99}'s claims, \citet{RCCA02} found no
such extremely violent variability events in a sample of 20 EGRET blazars
observed with $\sim 20$ min time-resolution along two or more nights each,
showing that the discrepancies between both works probably had their origin
in different methods for error control. In particular, \citet{RCCA02}
suggested that an inappropriate choice of the comparison and control stars
used for differential photometry could result in spurious fluctuations in
the differential light-curve.

Since it is desirable to firmly establish which is the real minimum
time-scale of blazar microvariability, we have undertaken an intensive
monitoring campaign focused on three particular objects with previous claims
of extremely violent flux variations, in order to check this behaviour
independently. We have made a careful analysis of the error sources involved
in the differential photometry following well-established procedures, thus
producing light-curves in which the significance of any variation is
quantitatively evaluated. This procedure allowed us to set on solid bases
the microvariability behaviour of blazars, showing that most claimed
extremely violent events, if not all, are very likely to be spurious results
produced by an inappropriate error handling. In Sect.~\ref{s_so} we give
details on the selected objects and the observations, while in
Sect.~\ref{s_da} we describe our methodology, with special attention to the
statistical error analysis. We show our results in
Sect.~\ref{s_resu}. Spurious variability results are exemplified in
Sect.~\ref{s_sv}.  We close in Sect.~\ref{s_sc} with some recommendations to
future observers.

\section{Sample and observations} \label{s_so}

In order to independently check the reported events of extremely violent
microvariability in blazars, we have selected three of the most variable
objects according to \citeauthor{XLZBL99}'s papers \citep{XLZBL99, XLB01,
XLZ02, XZD02, XZL04, DXL01}. Their names, equatorial coordinates, redshifts,
catalogued visual magnitudes, and classification are listed in columns 1 to
6 of Table~\ref{t_data}, respectively. Here follows a short description of
each blazar.

\paragraph*{\uno:} The optical spectrum  of this BL\,Lac shows very faint
emission lines, making its redshift determination rather uncertain
\citep{RS01}. A historical ($\sim 30$ years) $B$-band light-curve shows a
1.8 mag variation; in the visual band, flux changes up to $\Delta V \simeq
2.7$ mag have been recorded. A larger variation is reported at infrared
wavelengths, amounting to $\sim 6$ mag \citep{FL99}. On time-scales of
several months, \citet{FBBT93} reported a $\Delta V \simeq 0.9$ mag
variation.
\\ \citet{XZD02} reported variations up to $\Delta R =
0.32$ mag in 30 minutes (January 2001).

\paragraph*{\dos:}  This is another BL\,Lac whose redshift is still uncertain
\citep{FU00}. \citet{B80} reported variations up to $\sim 2$ mag in
its optical flux, over long time-scales. This object also displayed
fast polarization variability, both at optical ($P_\mathrm{opt} = 4 -
26\%$) and IR ($P_\mathrm{IR} = 4 - 19\%$) wavelengths \citep{PS80}. A
$V$-band light-curve compiled by \citet{FL00} shows a $\sim 3$ mag
change in 10 years, with smaller variations up to $\Delta V \simeq 1$
mag in about 1 year.
\\
\citet{BXL98} reported $\Delta B = 0.47$ mag in 22 minutes, while
\citet{XZL04} claimed $\Delta B = 0.56$ mag and $\Delta R = 0.66$ mag in
about 80 minutes.

\paragraph*{\tres:} This is a well-studied FSRQ, with a hard X-ray spectrum
\citep*{SSG97} and a powerful gamma-ray emission, detected by EGRET
\citep{TBD93, SBD96}. At radio frequencies, it has shown fast, large
amplitude flux changes \citep*{AAH81, AAH96}. Significant optical variations
were first reported by \citet{L72} over a $\sim 5$ years time-scale.  Its
historical light-curve since 1899 was reconstructed by \citet{LL75}; it
shows a long-term variation with a maximum range $\Delta B = 5.4$ mag,
including an outburst in 1948, after which the source brightness faded by
2.2 mag in 9 days.  \citet{GRSS00} report ``irregular variability of this
blazar on time-scales of days to weeks,'' with a $\Delta R \simeq 0.5$ mag
brightening in 84 days.\\
Extremely violent events, with the highest amplitudes and shortest
time-scales, were repeatedly claimed for this object: $\Delta R = 0.65$ mag
in 13 min \citep{XLB01}, $\Delta R = 2.0$ mag in 42 min \citep{DXL01}, and
$\Delta V = 1.68$ mag in 60 min \citep{XLZ02}.  A $\Delta R =1.35$ mag
``dip'' lasting $\Delta t=89$ minutes was reported by \citet{XZL04} in the
light-curve of this blazar.

\medskip

These three objects were the targets of our monitoring campaign, using the
2.15\,m ``Jorge Sahade'' telescope at CASLEO, Argentina, equipped with a
Roper--EEV $1340\times1300$ pixels CCD (gain: 2.3 electrons adu$^{-1}$;
read-out noise: 7.6 electrons). A focal-reducer provided a $\sim 9$ arcmin
diameter field, with a scale of 0.67 arcsec pix$^{-1}$.  The blazars \uno\
and \tres\ were followed during six consecutive nights in August 2004, while
\dos\ was observed along 5 nights (with a one-night gap, due to bad weather)
in January 2005. Atmospheric conditions were photometric
for $\sim 54\%$ of our observations, with some cirrus and/or bad seeing
during the remaining time. Moon illumination was always below 38\%
during the August 2004 run and below 69\%
during the January 2005 run.

Each object was repeatedly observed alternating between an
$R$ (Cousins) and a $V$ (Johnson) filter. Integration times ranged from 150
to 300\,s; thus, between 6 and 20 images of the same object and with the
same filter were obtained each night.

Science frames were bias-corrected and then flat-fielded using master bias
and flat-field frames (one flat for each filter) obtained by averaging 25
individual frames. The IRAF\footnote{%
IRAF is distributed by the National Optical Astronomy Observatories,
    which are operated by the Association of Universities for Research
    in Astronomy, Inc., under cooperative agreement with the National
    Science Foundation.}%
 package \textsf{ccdred} was used for this purpose.

\begin{table}
\caption{Selected objects.\label{t_data}}
\begin{tabular}{cccccc}
\hline
\hline
   Name  & $\alpha_{2000.0}$ & $\delta_{2000.0}$ & z & $m_V$ & Type \\
           & (h min s)         & ($^\circ ~~ ' ~~ ''$) & & (mag) &\\
 \hline 
 \uno  & $00~50~41.3$ & $-09~29~05$ & $0.22$ & $17.4$ & BL\,Lac \\
 \dos  & $07~57~06.6$ & $+09~56~35$ & $0.27$ & $14.5$ & BL\,Lac \\
 \tres & $15~12~50.3$ & $-09~06~00$ & $0.36$ & $16.5$ & HPQ \\
\hline
\end{tabular}
\end{table}


\section{Data analysis}
\label{s_da}

\subsection{Statistical error analysis}
\label{s_sea}

The usual technique to obtain light-curves for AGNs (and many other
astronomical objects) is differential photometry, i.e., the magnitude of the
target is measured against that of a comparison star recorded on the same
CCD frame. In this way, small fluctuations due to non-photometric conditions
are cancelled, since all star-like objects on the field of view are equally
affected \citep{HJ86}. A second star, measured against the same comparison
star, is often used as a stability check. In this way, variability is
evaluated by comparing the dispersions of two light-curves:
target--comparison star ($\sigma_\mathrm{T}$) on the one hand, and control
star--comparison star ($\sigma$) on the other. Assuming that the second
light-curve should only be affected by instrumental variations (i.e., both
stars are non-variable), a statistical criterion is often used
\citep[e.g.][]{JM97, RCC99} by introducing a parameter
$C=\sigma_\mathrm{T}/\sigma$ and requiring $C\geq 2.576$ for the source to
be considered as variable at a 99\% confidence level.

This technique is conceptually so simple that possible problems arising from
its misuse are sometimes looked-over, thus leading to the feeling that
differential photometry is almost immune to any error source. To prevent
this, several works have made useful recommendations to observers, based on
firm statistical bases. For example, \citet{H89} showed that very small
(optimum) apertures maximize S/N for point-source observations, when used
along with CCD growth curves to obtain instrumental magnitudes.  However, it
was later shown that such small photometric apertures should not be used for
the particular case of an AGN embedded in a bright host galaxy, since this
can lead to spurious variability results when the seeing FWHM changes along
the observations \citep*{CRC00}.

Regarding the correct choice of stars for comparison and control, a key work
is that by \citet*[hereafter HWM88]{HWM88}. These authors show that it is not
sufficient to simply select non-variable field stars for that purpose; these
stars should also closely match the target's magnitude (colour matching is
shown not to be so important). If not, the measured dispersion of the
target--comparison light-curve ($\sigma_\mathrm{T}$) will be different from
that of the control--comparison light-curve ($\sigma$), just from photon
statistics and other random-noise terms (sky, read-out noise), even in the
absence of any intrinsic variations in the target.

Since suitable stars are not always found (especially for high-Galactic
latitude fields), \citetalias{HWM88} give detailed calculations to derive a
corrective factor $\Gamma$ which properly scales $\sigma$ in order to match
the expected \emph{instrumental} dispersion $\sigma_\mathrm{T\,(INST)}$ of the
target--comparison light-curve. The computation of $\Gamma$ thus requires
the knowledge of the relevant CCD parameters, as well as mean values of the
sky brightness and magnitudes of target and stars. Following
\citetalias{HWM88} (their eq.\ 13) this corrective factor can be written as
\begin{equation}
\Gamma^2 = \left( {N_{\mathrm{S}2} \over N_\mathrm{T}}\right)^2 \left[
  {N_{\mathrm{S}1}^2 (N_\mathrm{T}+P) +N_\mathrm{T}^2 (N_{\mathrm{S}1}+P)
  \over N_{\mathrm{S}2}^2 (N_\mathrm{T}+P) +N_\mathrm{T}^2
  (N_{\mathrm{S}2}+P) } \right]\, ,
\end{equation}
where $N$ stands for total (sky-subtracted) counts within the aperture,
while sub-indices T, S1, and S2 correspond to the target, comparison star,
and control star, respectively. The factor $P$ takes into account
common noise-terms, being $P = n_\mathrm{pix} (N_\mathrm{sky} +
N_\mathrm{RON}^2)$, where $n_\mathrm{pix}$ is the number of pixels within
the aperture, $N_\mathrm{sky}$ is the sky level, and $N_\mathrm{RON}$
is the read-out noise. Median values are used for objects and sky.

Thus, using the scaled $\sigma$, the confidence parameter is now re-written
 as the quotient between the observed target--comparison light-curve
 dispersion and its expected dispersion just from instrumental and
 photometric errors:

\begin{equation}
{C \over \Gamma} = {\sigma_\mathrm{T} \over \Gamma \; \sigma} =
{\sigma_\mathrm{T} \over \sigma_\mathrm{T\,(INST)}}\, .
\end{equation}
In the ideal case when all three objects are of the same magnitude, then
$\Gamma =1$, and the original definition of $C$ is recovered.

Many AGN variability studies, although not explicitly applying
\citetalias{HWM88}'s method, use comparison and control stars with apparent
magnitudes very close to that of the target object \citep[e.g.,][]{RCC99,
SSGW04}, thus ensuring $\Gamma \simeq 1$. Their results are in this way
trustworthy, since the variability confidence levels of all light-curves are
properly estimated.  However, it is a fact that, whenever extremely violent
variations have been claimed, they resulted from differential photometry
using comparison and control stars more than $\sim 2$ mag (and up to $\sim
5$ mag) brighter than the AGN, and without any dispersion scaling. This
flawed procedure leads to a severe overestimation of the confidence
parameter. In what follows, we will further illustrate this point with
results from our own observations.


\subsection{Photometry}

We used the IRAF package \textsf{apphot} to obtain aperture photometry for
the three blazars and several isolated, non-saturated stars in each
field. Aperture radii were set at 8 pix ($5\farcs4$, i.e., between $\sim
1.5$ and 2 times the seeing FWHM) in order to prevent against any unwanted
effect due to light from the host galaxy under varying seeing conditions
\citep{CRC00}. We then selected the two most suitable stars in each field to
be used as comparison (S1) and control (S2), by requiring them to be
non-variable and with magnitudes as close as possible to that of the
blazar. In fact, the best results are obtained when S1 is slightly brighter
than the target \citepalias{HWM88}, so, we tried to fulfil this condition,
too.

\begin{figure*}
\fbox{\includegraphics[width=0.3\hsize]{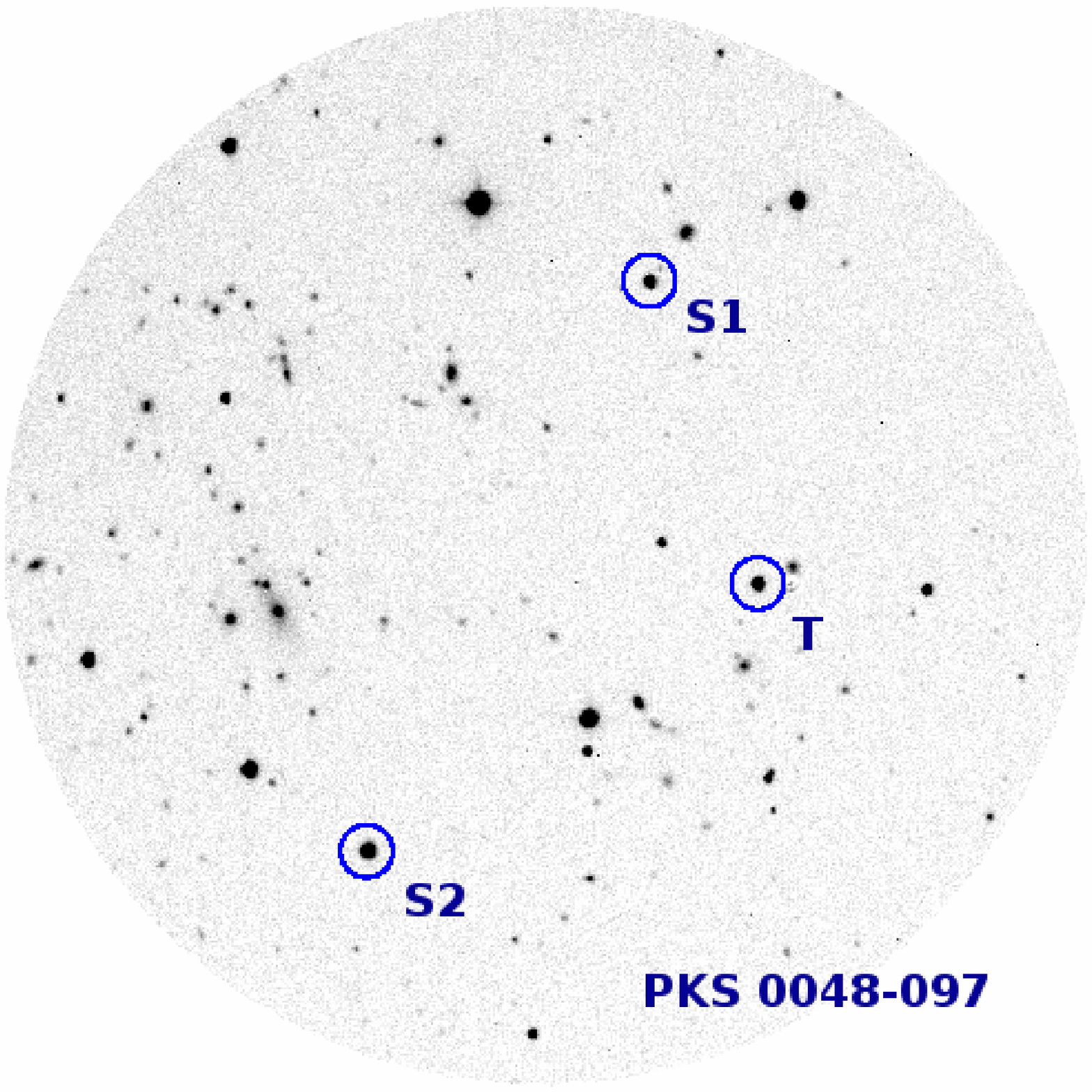}}
\fbox{\includegraphics[width=0.3\hsize]{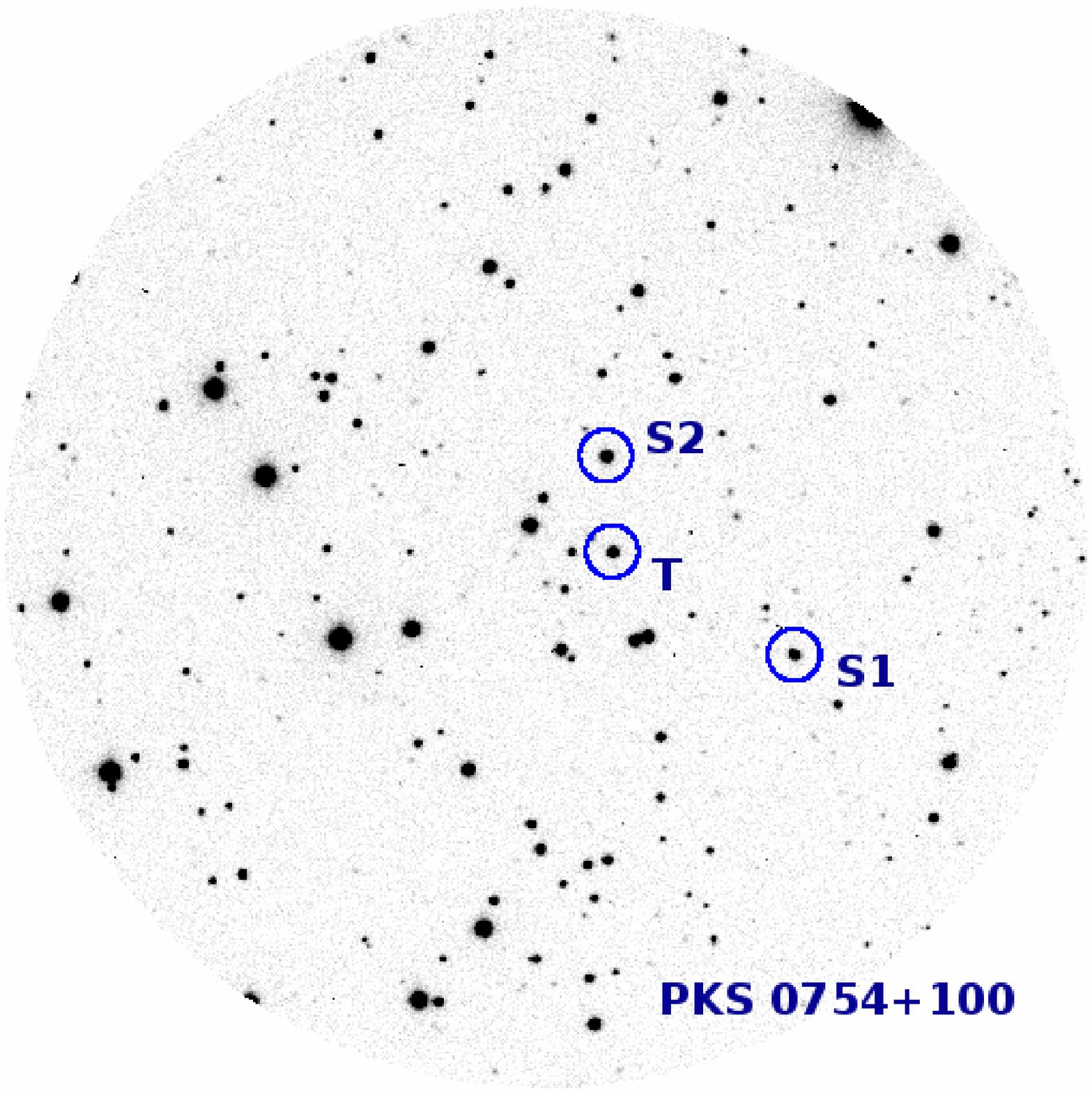}}
\fbox{\includegraphics[width=0.3\hsize]{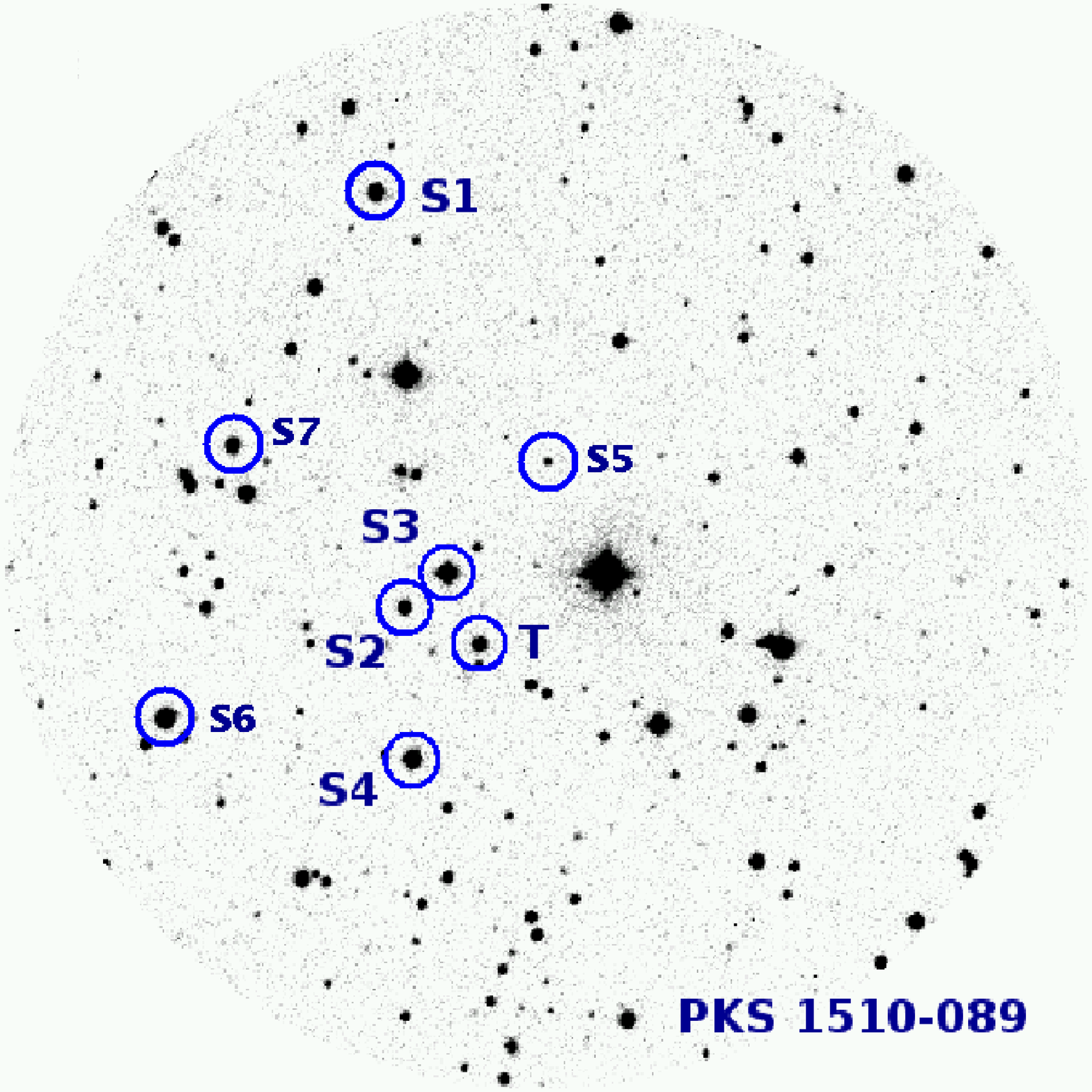}}
\caption{Finding charts for the observed fields. Each one is $\sim 9$
arcmin across, with North up, and East, to the left. Our target blazars
are marked with $T$, while S$_i$ are the stars used for differential
photometry (see text). \label{f_cartas}}
\end{figure*}

We show finding charts for the three fields in Fig.~\ref{f_cartas}. Note
that we generally did not use known \emph{standard} stars in the blazars'
fields, since most of these are too bright.  The exceptions are star S2 in the
field of \uno, which is star \#4 in \citet{VRL98}, and star S2 in the
field of \dos, which is star $D$ in \citet*{MMM83}.

After constructing the differential light-curves, they were checked for any
suspicious data-points, such as sudden changes in the control light-curve,
and/or in only one of the photometric bands. A few of these events, due to
cosmic-ray hits, were found and corrected. In the next section, we discuss
our results.

\section{Results}
\label{s_resu}

\subsection{Inter-night results}

\begin{table*}
\caption{Inter-night results \label{t_todo}}
\begin{center}
\begin{tabular}{ccccccccc}
\hline
Object  & Filter & UT Date  & $\sigma$ & $\Delta t$ & $C/ \Gamma$ &
Variable? & $\Delta m_{\rm{max}}$ & Nr. \\
 {} & {} & (dd--dd/mm/yy) & [mag] & [h] & {} & {} & [mag] & {}\\
\hline
\hline
\uno  & $R$ & 08--13/08/04 & 0.007 & 122.50 & 13.61 & yes & 0.31 & 53\\
      & $V$ &              & 0.007 & 122.50 & 13.07 & yes & 0.29 & 54\\
\dos  & $R$ & 14--19/01/05 & 0.008 & 125.21 & 10.18 & yes & 0.36 & 74\\
      & $V$ &              & 0.007 & 124.94 & 12.97 & yes & 0.42 & 74\\
\tres & $R$ & 08--13/08/04 & 0.006 & 122.24 & 31.89 & yes & 0.61 & 55\\
      & $V$ &              & 0.007 & 122.28 & 23.99 & yes & 0.57 & 58\\
\hline
\end{tabular}
\end{center}
\end{table*}

\begin{figure}
\includegraphics[width=0.9\hsize]{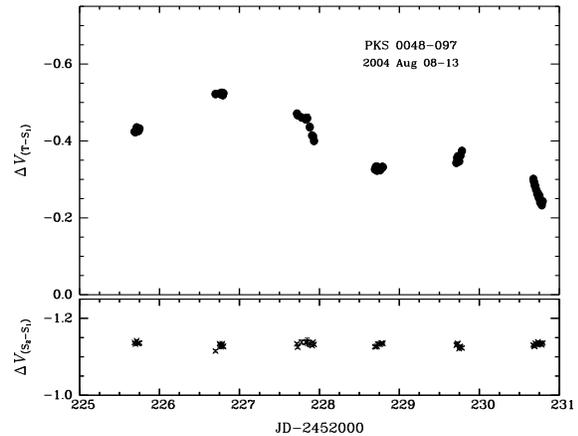}
\caption{Differential $V$ band light-curve for \uno, whole campaign. Upper
panel: AGN $-$S$_1$ ; lower panel: S$_2-$S$_1$. \label{f_0048Vtodo}}
\end{figure}

\begin{figure}
\includegraphics[width=0.9\hsize]{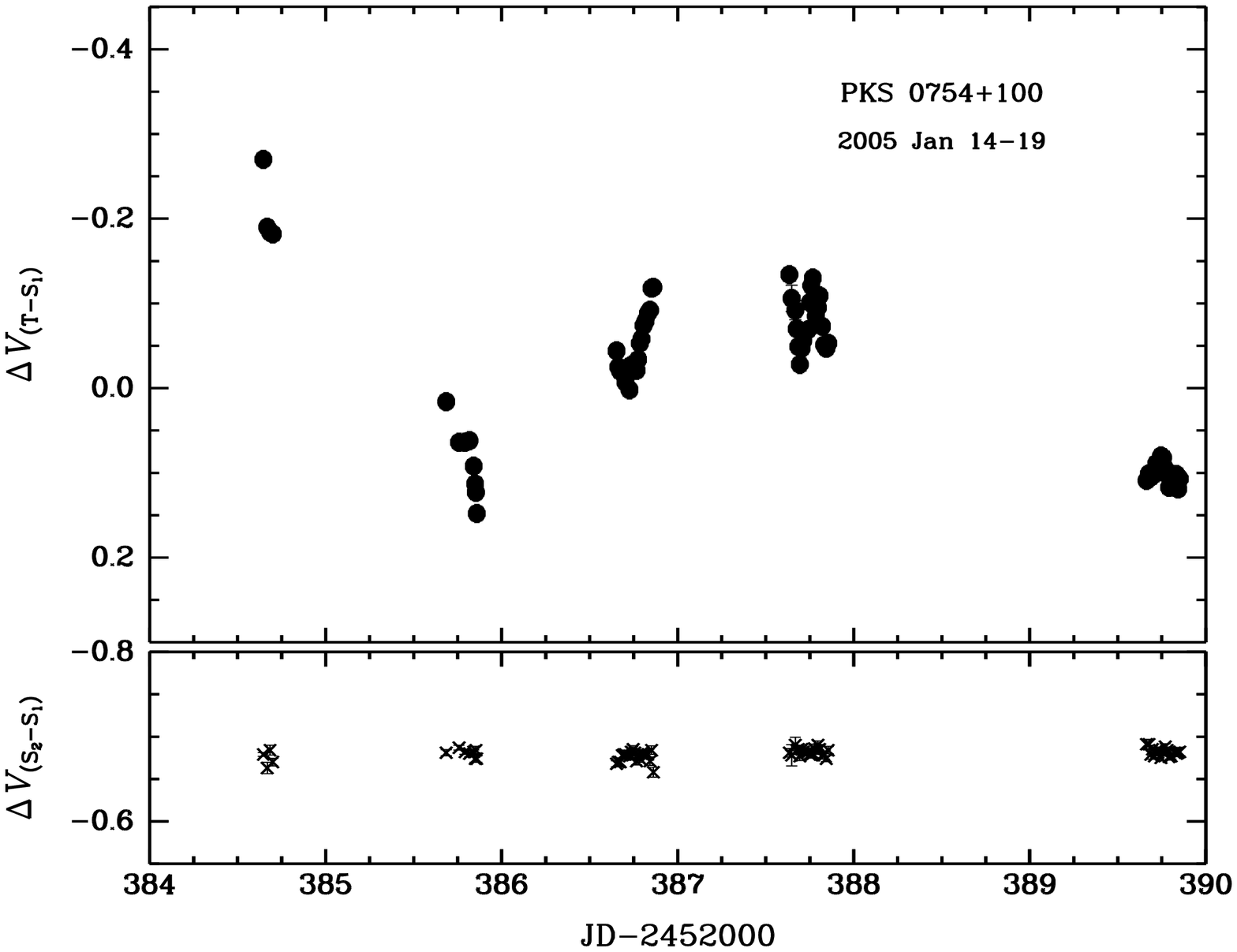}
\caption{Same as Fig.~\ref{f_0048Vtodo} for \dos. \label{f_0754Vtodo}}
\end{figure}

\begin{figure}
\includegraphics[width=0.9\hsize]{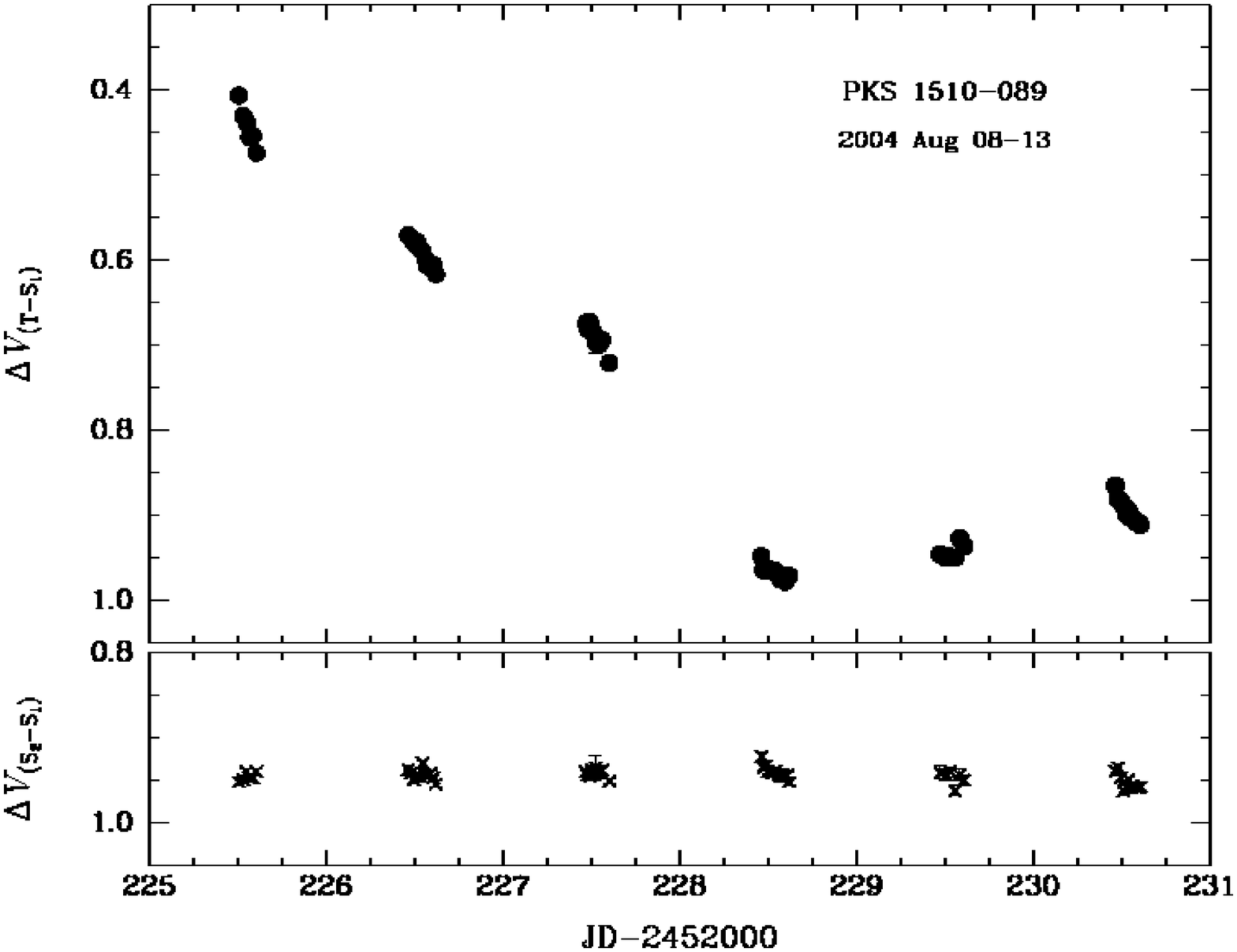}
\caption{Same as Fig.~\ref{f_0048Vtodo} for \tres. \label{f_1510Vtodo}}
\end{figure}

Results from the whole campaign are given in Table~\ref{t_todo}, where
columns 1 to 6 are, respectively: object name, filter, dates of observation,
control--comparison light-curve dispersion ($\sigma$), total time between
the first and last data points ($\Delta t$), and scaled confidence parameter
($C/ \Gamma$). In column 7 we state whether the blazar was variable or not
during our observations, according to the adopted criterion. The last two
columns give, respectively, the maximum variation amplitude (in magnitudes)
along the whole campaign, and the total number of data points in each
light-curve.  All light-curves are graphically shown in
Figs.~\ref{f_0048Vtodo}--\ref{f_1510Vtodo}. Each figure shows the
target--comparison light-curve in the upper panel, and the
control--comparison light-curve in the lower panel against heliocentric
Julian Date. Note that the vertical axis scale is always the same, in both
panels of all three figures.

It is evident that the three blazars were variable, with very high
statistical significances, at these inter-night time-scales. Maximum
amplitudes reached about half a magnitude for \tres, and somewhat smaller
values for the other two objects. Note that there is a good agreement
between variability parameters in the $V$ and $R$ bands.

Individual error bars and control--comparison dispersions are larger for a
few nights, when scattered moonlight and/or tracking errors affected our
photometry. However, this had no evident impact on our ability to assess the
object's variability.  On the other hand, the data corresponding to the
fifth observing night for \tres\ are affected by a $0.065$ mag zero-point
shift, due to technical problems. We corrected the graph in
Fig.~\ref{f_1510Vtodo} for this effect, but we did not consider those data
for the inter-night analysis.

\subsection{Intra-night results}

\begin{table}
\caption{Intra-night results for \uno \label{t_uno}}
\begin{center}
\begin{tabular}{ccccccr}
\hline
UT Date  & Filter &  $\sigma$ & $\Delta t$ & Variable? & $C/ \Gamma$ & Nr.\\
 (dd/mm/yy) & {} & [mag] & [h] & {} & {} & {}\\
\hline
\hline
08/08/04  & $R$ & 0.004 & 1.44 & no  & 1.35 & 6\\
          & $V$ & 0.008 & 1.44 & no  & 0.76 & 7\\
09/08/04  & $R$ & 0.005 & 2.33 & no  & 1.37 & 8\\
          & $V$ & 0.007 & 2.38 & no  & 0.27 & 8\\
10/08/04  & $R$ & 0.008 & 5.11 & yes & 3.22 & 10\\
          & $V$ & 0.009 & 5.11 & yes & 2.89 & 10\\
11/08/04  & $R$ & 0.007 & 2.38 & no  & 0.59 & 8\\
          & $V$ & 0.005 & 2.38 & no  & 0.80 & 9\\
12/08/04  & $R$ & 0.006 & 1.66 & no  & 2.56 & 9\\
          & $V$ & 0.006 & 1.66 & no  & 1.71 & 7\\
13/08/04  & $R$ & 0.004 & 2.83 & yes & 4.82 & 12\\
          & $V$ & 0.004 & 2.78 & yes & 5.54 & 13\\
\hline
\end{tabular}
\end{center}
\end{table}


\begin{table}
\caption{Intra-night results for \dos \label{t_dos}}
\begin{center}
\begin{tabular}{ccccccr}
\hline
UT Date  & Filter &  $\sigma$ & $\Delta t$ & Variable? & $C/ \Gamma$ & Nr.\\
 (dd/mm/yy) & {} & [mag] & [hs] & {} & {} & {}\\
\hline
\hline
14/01/05  & $R$ & 0.004 & 1.82 & yes & 9.66 & 8\\
          & $V$ & 0.011 & 1.27 & yes & 4.02 & 8\\
15/01/05  & $R$ & 0.007 & 3.91 & yes & 5.78 & 10\\
          & $V$ & 0.006 & 3.91 & yes & 7.23 & 8\\
16/01/05  & $R$ & 0.007 & 4.75 & yes & 4.24 & 19\\
          & $V$ & 0.008 & 5.02 & yes & 4.58 & 20\\
17/01/05  & $R$ & 0.010 & 5.33 & yes & 3.01 & 20\\
          & $V$ & 0.005 & 5.33 & yes & 6.11 & 20\\
19/01/05  & $R$ & 0.007 & 4.56 & no  & 1.94 & 17\\
          & $V$ & 0.006 & 4.51 & no  & 1.81 & 18\\
\hline
\end{tabular}
\end{center}
\end{table}

\begin{table}
\caption{Intra-night results for \tres \label{t_tres}}
\begin{center}
\begin{tabular}{ccccccr}
\hline
UT Date  & Filter &  $\sigma$ & $\Delta t$ & Variable? & $C/ \Gamma$ & Nr.\\
 (dd/mm/yy) & {} & [mag] & [hs] & {} & {} & {}\\
\hline
\hline
08/08/04  & $R$ & 0.002  & 2.38 & yes & 12.88 & 6\\
          & $V$ & 0.004  & 2.45 & yes & 6.68 & 6\\
09/08/04  & $R$ & 0.004  & 3.74 & yes & 3.38 & 12\\
          & $V$ & 0.005  & 3.74 & yes & 3.15 & 12\\
10/08/04  & $R$ & 0.006  & 3.17 & yes & 3.91 & 10\\
          & $V$ & 0.004  & 3.17 & yes & 3.75 & 10\\
11/08/04  & $R$ & 0.004  & 2.38 & no  & 2.16 & 11\\
          & $V$ & 0.008  & 3.77 & no  & 1.01 & 13\\
12/08/04  & $R$ & 0.013  & 3.17 & no  & 1.10 & 6\\
          & $V$ & 0.008  & 3.24 & no  & 1.16 & 6\\
13/08/04  & $R$ & 0.005  & 3.38 & no  & 2.42 & 10\\
          & $V$ & 0.009  & 3.43 & no  & 1.62 & 11\\
\hline
\end{tabular}
\end{center}
\end{table}

\begin{figure}
\includegraphics[width=0.9\hsize]{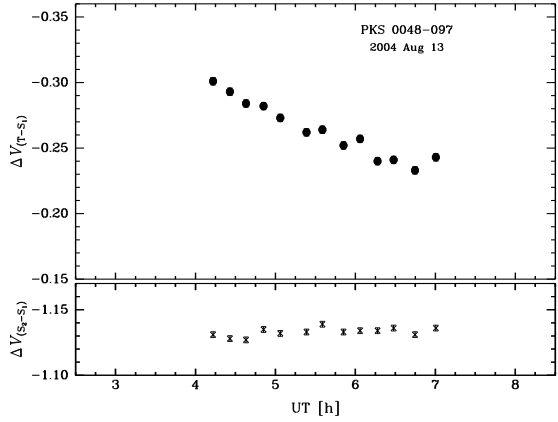}
\caption{Differential $V$ band light-curve for \uno, on the night of Aug 13,
2004. Upper panel: AGN $-$S$_1$ ; lower panel:
S$_2-$S$_1$. \label{f_0048Vaug13}}
\end{figure}

\begin{figure}
\includegraphics[width=0.9\hsize]{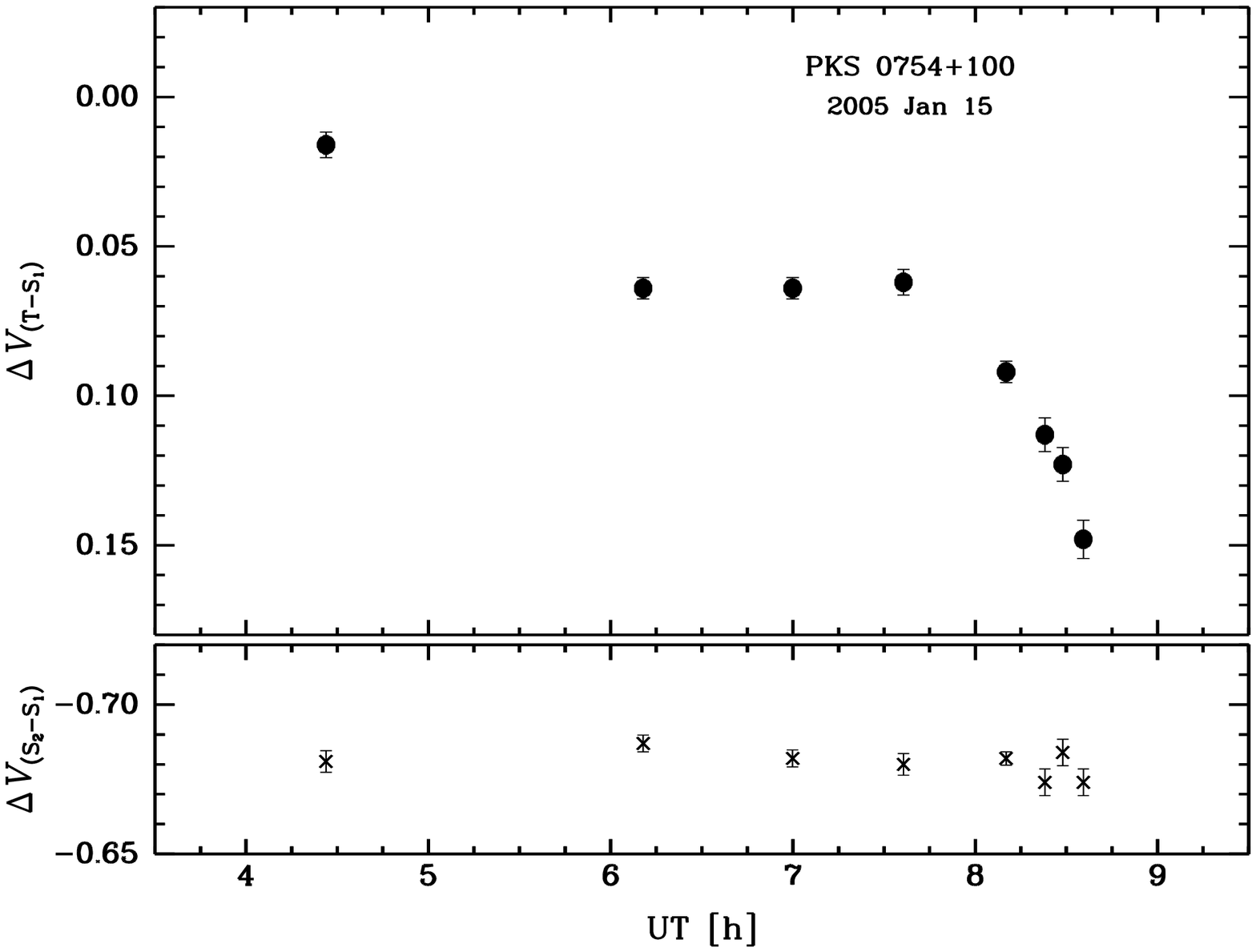}
\caption{Differential $V$ band light-curve for \dos, on the night of Jan 15,
2005. Upper panel: AGN $-$S$_1$ ; lower panel:
S$_2-$S$_1$. \label{f_0754Vjan15}}
\end{figure}

\begin{figure}
\includegraphics[width=0.9\hsize]{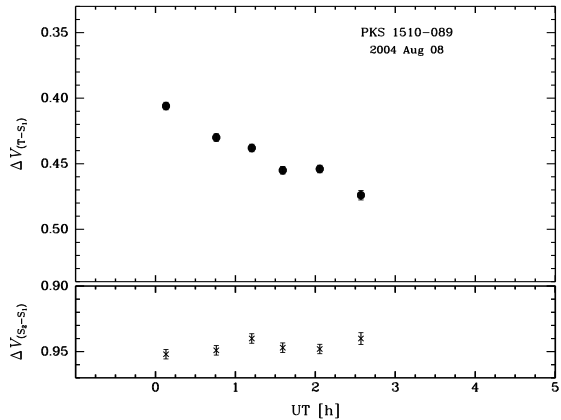}
\caption{Differential $V$ band light-curve for \tres, on the night of Aug 8,
2004. Upper panel: AGN $-$S$_1$ ; lower panel:
S$_2-$S$_1$. \label{f_1510Vaug08}}
\end{figure}

Tables~\ref{t_uno}, \ref{t_dos}, and \ref{t_tres} summarise the intra-night
variability results for our three targets. Columns 1 to 6 give date of
observation, filter, control--comparison light-curve dispersion ($\sigma$),
time spanned by the observations ($\Delta t$), and scaled confidence
parameter ($C/ \Gamma$), respectively. Column 5 states whether the blazar
was considered to be variable or not during each night. The last column
gives the number of data points.  We also show, for each blazar, the $V$
light-curve for the night when the largest variation was detected
(Figs.~\ref{f_0048Vaug13}--\ref{f_1510Vaug08}).

All three blazars displayed microvariability, with amplitudes up to $\Delta m
\simeq 0.08$ mag in 1 hour; however, each of them was classified as
non-variable for at least one night. This does not mean that, in such cases,
the object's flux was completely constant; all we can say is that any
possible variation was then below our confidence threshold.

It is thus clear that no extremely violent behaviour was detected in any
source along our whole campaign. The statistical significance of this result
can be assessed as follows. Let us define the \emph{duty cycle} ($DC$) for
extremely violent microvariability as the fraction of the observing time for
which the object displayed large amplitude ($\Delta m \gtrsim 0.5$ mag),
fast ($\Delta t \lesssim 45$ min) flux changes. From the works by Xie's
group (see Sect.~\ref{s_intro}) a duty cycle $DC \simeq 50\%$ is inferred;
if we accept this number, the fact that we did not detect any such extremely
violent event along 17 observing sessions would have a very low probability
($\sim 10^{-5}$).

This result strongly implies that extremely violent microvariability reports
should be carefully analysed, disentangling real flux changes in the source
from systematic errors. In the next section we show how spurious results can
be obtained from wrong error handling.


\section{Spurious variability}
\label{s_sv}

As an illustrative example of the effects of a bad choice of comparison and
control stars, let us consider our data for \tres\ on the night of Aug.\ 13,
2004. Our originally selected stars approximately follow the prescriptions
given in Sect.~\ref{s_sea}: $\Delta V (\mathrm{T}-\mathrm{S}_1) = 0.90$,
$\Delta V (\mathrm{S}_2-\mathrm{S}_1) = 0.96$; i.e., comparison star not
much brighter than target AGN and control star slightly fainter than the
AGN. With these stars, we obtain a confidence parameter $C=1.55$, and a
scaled confidence parameter ${C \over \Gamma}=1.62$; i.e., the blazar
is classified non-variable in both cases. We now use two significantly brighter
stars, shown as S$_3$ and S$_4$ in Fig.~\ref{f_cartas}: $\Delta V
(\mathrm{T}-\mathrm{S}_3) = 1.95$, $\Delta V (\mathrm{S}_4-\mathrm{S}_3) =
0.58$ (S$_3$, S$_4\equiv$ star~\#4, star~\#6 in \citealt{RVL98} $\equiv$
star A, star B in \citealt{VRG97}, respectively). This choice gives a
confidence parameter $C=3.72$, implying that the blazar should be considered
as variable during that night. However, after correcting for the large flux
difference between the objects, a scaled confidence parameter ${C \over
\Gamma}=1.32$ is obtained, thus classifying the target as non-variable.

\begin{figure}
\includegraphics[width=0.9\hsize]{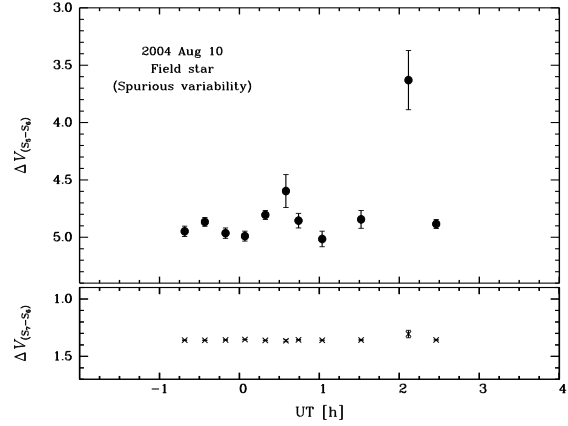}
\caption{Differential $V$ band light-curve for a faint non-variable star in
the field of \tres, on the night of Aug 10, 2004. Upper panel: S$_5$-S$_6$;
lower panel: S$_7-$S$_6$. The apparent extreme variability is just a
spurious result.
\label{f_1510Vespur}}
\end{figure}

The preceding analysis shows that, although a certain level of intrinsic
variability may be present in the target, its significance is severely
overestimated when stars $\sim 2$ mag brighter than the target are chosen
for comparison and control. We now want to test if dramatic, although
spurious, variability events may be produced when still brighter stars are
used. Since any star more than $\sim 3$ mag brighter than our targets was
always saturated on our images, we selected a faint, non-variable star in
the field of \tres\ to illustrate this effect. This star is labelled as
S$_5$ in Fig.~\ref{f_cartas}, and we used stars S$_6$ and S$_7$ as
comparison and control, respectively (S$_6 \equiv$ star~\#5 in
\citealt{RVL98}). Hence, we reproduce a situation where the target is $\sim
5$ mag fainter than the comparison star, while this last, in turn, is $\sim
1.4$ mag brighter than the control star.  The result of this completely
inappropriate choice of stars is shown in Fig.~\ref{f_1510Vespur}: in about
one hour, the source seems to brighten by $\sim 0.4$ mag, then returning to
its original flux level; later, a spectacular ``outburst'' brightens the
object by $\sim 1.2$ mag in about 35 minutes. Note the stability of the
control -- comparison light-curve; without any dispersion scaling, the
variability confidence parameter is $C = 24.0$, thus giving the impression
of a highly significant flux variation. However, the scaled confidence
parameter is just ${C \over \Gamma} = 1.0$, clearly establishing that no
significant variability is present in the data, with any fluctuation in the
object -- comparison light-curve being the result of photometric errors
differently affecting stars of very disparate magnitudes. In this particular
case, both ``outbursts'' coincide with sudden changes in atmospheric
transparency, due to passing cirrus. Several factors contribute to make
errors larger during such events: a lower flux level from the star, higher
sky level and its associated rms, changes in the seeing FWHM, etc.  Brighter
stars are relatively less affected by these effects than the faint target,
thus leading to the apparent variations in the differential
light-curve. Under different observational conditions and with different
photometric techniques, it is likely that spurious ``dips'' instead of
``outbursts'' may be produced in the light-curve.

These results underscore the absolute necessity of using the method
described in \citetalias{HWM88} whenever suitable comparison and control
stars cannot be found. Not doing so will very probably lead to spectacular,
although completely false, results.

In some cases, extremely violent variability events have been claimed to be
periodic \citep{XZL04}. This is not surprising, since the observations are
periodically repeated, and, if the same inadequate photometric techniques
are used, similar spurious results will be obtained. These systematic errors
have generated some completely ill-motivated theoretical models
\citep{WZP05}.

\section{Summary and conclusions}
\label{s_sc}

Repeated claims for the detection of extremely violent optical variability
in blazars have been raised up in recent years \citep{XLZBL99, XLB01, XLZ02,
XZD02, XZL04, DXL01}. These claimed events are characterised by fast flux
changes $\Delta m \gtrsim 0.5$ mag in a few tens of minutes, and reaching up
to, for example, a 2 mag variation in $\sim40$ min reported for \tres.
However, other studies have found that the typical minimum time-scale for
such large-amplitude variations in blazars is of several hours or still
larger \citep[e.g.,][]{RCCA02}.  We have thus undertaken an observational
campaign, targeting the blazars \uno, \dos, and \tres, devised to shed light
on this controversy. This paper presents its results, showing that, although
microvariability was clearly detected in our three targets, no extremely
violent optical variability event was detected along 110 hours of
observation. The largest fast flux variations we detected, instead, amount to
$\lesssim 0.1$ mag in about one hour.

We show that this discrepancy is most likely due to systematic errors
introduced during the observations and photometry. In particular, the use
(without any correction) of stars much brighter than the target for
differential photometry, directly leads to an overestimation of the
significance of any detected variability. Moreover, under certain specific
conditions, it easily gives place to spurious variability closely
reproducing extremely violent microvariability events as those reported in
the papers by Xie and coworkers.

The following  recommendations should thus be followed in order to prevent
against spurious variability results:

The target object must be neither underexposed, nor saturated on all science
frames.  Differential light-curves should be made using comparison and
control stars as close in magnitude as possible to the target. Published
standard stars in blazar fields are usually too bright for this purpose;
they should be used just for calibration to the standard system, through a
few short exposure-time frames.

If no suitable stars can be selected for differential photometry, brighter
(or fainter) stars may be used, provided that the variance of the
control--comparison light-curve is properly scaled. This should be done
following the method presented in \citetalias{HWM88} (in fact, it is
\emph{always} recommended to use \citetalias{HWM88} method).

Any remarkable flux change should be critically verified, looking for
cosmic-ray hits, sudden changes in atmospheric transparency, or any
instrumental effect (see also \citealp{CRC00} for spurious variability
induced by seeing FWHM changes).

We conclude by saying that a critical evaluation of past and future claims for
extremely violent microvariability events in blazars is needed before any
radical revision of blazar models be required.

\section*{Acknowledgments}

This work received financial support from ANPCyT (PICT\,03-13291
BID\,1728/OC-AR) and CONICET (PIP 5375). Additional support from IALP
(UNLP--CONICET) is also acknowledged. We thank Cecilia Fari\~na for help
with the observations, and CASLEO staff members A. de Franceschi,
R. Jakowczyk, and P. Ostrov for their skillful assistance at the telescope.
We wish to thank the anonymous referee for his/her very constructive comments.



\end{document}